\documentclass[prl,aps,groupedaddress,twocolumn]{revtex4}
\usepackage{amssymb,amsmath,amsfonts,epsfig,graphicx,latexsym,bm,color}
\usepackage{soul}

\begin{document}
\title{Observing Chiral Superfluid Order by Matter-Wave Interference}
\author{T. Kock$^{1,*}$, M. \"{O}lschl\"{a}ger$^{1,}$\footnote{These authors contributed equally to this work.}, A. Ewerbeck$^{1}$, W.-M. Huang $^{1,2}$, L. Mathey$^{1,2}$, and A. Hemmerich$^{1,2,}$\footnote{e-mail: hemmerich@physnet.uni-hamburg.de}}
\affiliation{$^{1}$Institut f\"{u}r Laser-Physik, Universit\"{a}t Hamburg, Luruper Chaussee 149, 22761 Hamburg, Germany}
\affiliation{$^{2}$The Hamburg Centre for Ultrafast Imaging, Luruper Chaussee 149, 22761 Hamburg, Germany}
\date{\today}

\begin{abstract}
The breaking of time reversal symmetry via the spontaneous formation of chiral order is ubiquitous in nature. Here, we present an unambiguous demonstration of this phenomenon for atoms Bose-Einstein condensed in the second Bloch band of an optical lattice. As a key tool we use a matter wave interference technique, which lets us directly observe the phase properties of the superfluid order parameter and allows us to reconstruct the spatial geometry of certain low energy excitations, associated with the formation of domains of different chirality. Our work marks a new era of optical lattices where orbital degrees of freedom play an essential role for the formation of exotic quantum matter, similarly as in electronic systems.
\end{abstract}


\maketitle
The determination of the symmetry properties of superfluid or superconducting states requires phase information, which is often difficult to obtain in electronic condensed matter. In contrast, ultracold atomic gases provide a straight forward technique to study phase coherence by interference \cite{And:97, Had:06, Hof:07, Gro:11}, similarly as in the daily practice in optical interferometry. Quantum degenerate atomic gases filled into periodic potentials made of laser light establish an extremely versatile and well controlled experimental platform for studying superfluid order \cite{Lew:07, Blo:08}. Prospects to form and probe novel unconventional superfluids in such optical lattices have caused widespread excitement, as is documented by a wealth of recent theory proposals \cite{Liu:06, Wu:09, Li:11, Mar:12, Liu:13, Li:12, Cai:11, Cai:12, Sun:12, Pin:13, Heb:13, Li:14}. On the experimental side Bose-Einstein condensates were recently formed in higher bands of optical lattices, where orbital degrees of freedom lead to degeneracies of the states with minimal energy \cite{Wir:11, Oel:11, Oel:12}. The second Bloch band of a bipartite optical square lattice is a basic paradigm. It possesses two inequivalent energy minima at the boundary of the first Brillouin zone, which can be tuned to be energetically degenerate \cite{Wir:11, Oel:13}. It was shown in Ref.~\cite{Wir:11} that bosonic atoms loaded to the second band condense to each of these minima in equal shares. Furthermore, upon the assumption that these two condensates possess a fixed relative phase $\phi$, such that the two condensation points equally contribute to the formation of a single condensate, it was shown in Ref.~\cite{Oel:13}, that interaction constrains this relative phase to the two possible values $\phi = \pm \pi/2$. Hence, the formed condensate wave function locally exhibits chiral order in position space with the consequence of broken time-reversal symmetry. This form of chiral order in configuration space has a momentum space analogy in the global chiral symmetry of the superfluid order parameter proposed for the Cooper pairs of certain electronic superconductors \cite{Mae:94}.

\begin{figure}
\includegraphics[scale=1, angle=0, origin=c]{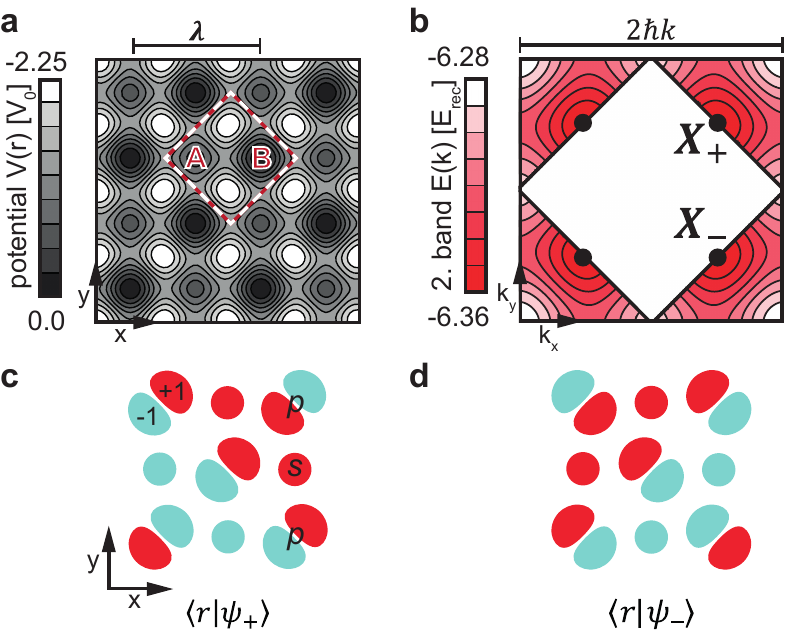}
\caption{\label{Fig.1} (color online). (a) Bipartite lattice geometry. The unit cell with two potential minima denoted by A and B is highlighted. $\lambda = 1064\,$nm. (b) The second band is plotted across the second Brillouin zone. Two degenerate band minima in the points $X_{\pm}$ arise. (c) and (d) The spatial wave functions associated with the Bloch states $|\psi_{\pm} \rangle$ associated with $X_{\pm}$ are composed of local $s$- and $p$-orbitals with the local phases $\pm 1$ (indicated by red and green color) arranged in order to maximize tunneling.}
\label{fig:1}
\end{figure}

In this work we provide the missing link for an unambiguous demonstration of chiral superfluid order in the second band of a bipartite optical square lattice by proving phase locking between both condensation points with a matter wave interference technique. By introducing a potential barrier, two entirely independent atomic samples are produced in the second band in well separated spatial regions of the lattice, which are subsequently brought to interference. Analysis of the emerging interference patterns reveals that in each of the two regions the relative phase between the two condensation points is locked to a fixed value modulo $\pi$, which corresponds to the spontaneous breaking of the expected $Z_2$ symmetry, associated with the two possible signs of chirality. By comparing the interference patterns with calculations \cite{Pol:06}, we can identify low energy excitations, arising at increased temperatures, associated with the formation of domains with different chirality. 

\begin{figure}
\includegraphics[scale=0.95, angle=0, origin=c]{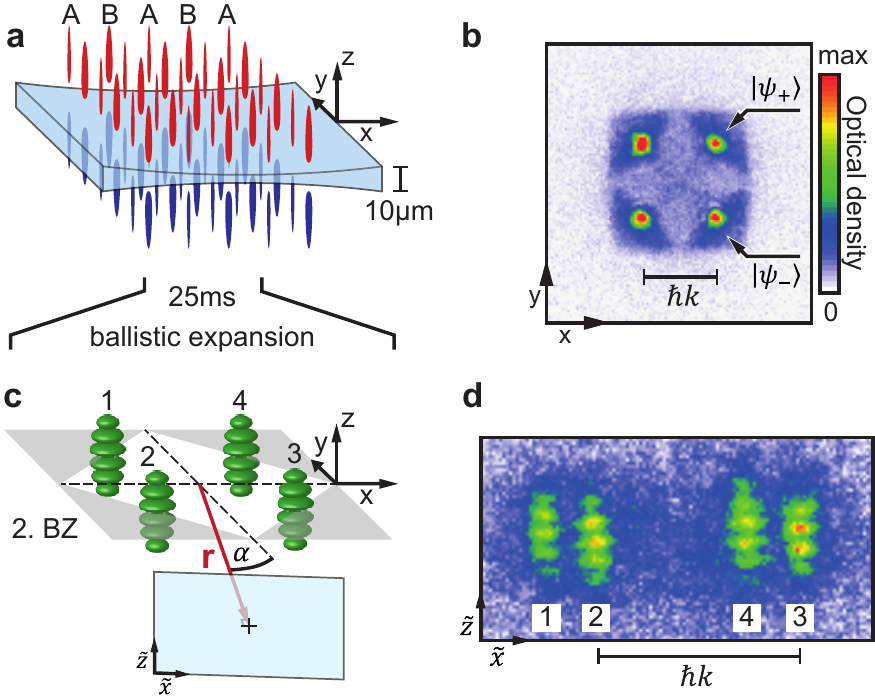}
\caption{\label{Fig.2} (color online). (a) The tubular minima of the lattice potential are split into upper (red) and lower (blue) sections with a cylindrically focused laser beam (light blue). Condensates are simultaneously prepared in the $X_\pm$-points of the second band in the two regions of the lattice spatially separated by the light sheet. (b) Band mapping image, showing equally populated condensates at the minima of the second band $X_{\pm}$. (c) Sketch of atomic spatial distribution after a ballistic expansion of $25\,$ms. The grey area corresponds to the second Brillouin zone. Each of the four Bragg maxima 1-4, corresponding to the $X_\pm$-points in Fig.~\ref{fig:1}(b), carries an interference grating aligned with the $z$-axis. (d) Absorption image along the line of sight indicated by the vector $\mathbf{r}$ in (c), which lies in the $xy$-plane and encloses an angle of $\alpha=13^\circ$ with the $y$-axis.}
\label{fig:2}
\end{figure}

We employ the optical chequerboard lattice potential, also used in Refs.~\cite{Wir:11, Oel:11, Oel:12, Oel:13}, which extends in the $xy$-plane, with the $z$-direction confined by a harmonic trap. Fig.~\ref{fig:1}(a) summarizes the lattice geometry with a unit cell comprised of two classes of lattice sites denoted as A and B. The difference in the potential depth of A- and B-sites $\Delta V = V_A-V_B$ can be tuned. Fig.~\ref{fig:1}(b) shows the second Bloch band in the extended Brillouin zone scheme. It provides two energy minima, denoted as $X_\pm$, which are located at the two inequivalent quasi-momenta $\frac{1}{2}\hbar k (1,\pm 1)$ at the edges between the first and second Brillouin zones with $k \equiv 2 \pi/ \lambda$ and $\lambda = 1064\,$nm the wavelength of the lattice beams. The bipartite lattice geometry gives rise to hybridized Bloch bands, i.e. the Bloch states $|\psi_{\pm} \rangle$ associated with $X_\pm$ are composed of local $s$-orbitals in the shallow wells and local $p$-orbitals in the deep wells as is shown in Fig.~\ref{fig:1}(c) and (d). Minimization of the kinetic energy imposes the local phases $\psi_{\pm}(r)/|\psi_{\pm}(r)|$ with $\psi_{\pm}(r)\equiv \langle r | \psi_{\pm} \rangle$ to match for adjacent orbital lobes such that a striped pattern $\psi_{\pm}/|\psi_{\pm}| = \pm 1$ arises. Hence, $\psi_{\pm}(r)$  represent standing matter waves along the diagonal directions $x\pm y$ in Fig.~\ref{fig:1}(a). In addition to the lattice potential, by adding a laser beam with 763~nm wavelength propagating along the $x$-direction with a tight cylindrical focus with respect to the $z$-direction, we introduce a potential barrier with a height of $2\,E_\textrm{rec}$ and $d_z \approx 10\,\mu$m thickness in the $z$-direction as illustrated in Fig.~\ref{fig:2}(a). This acts to split the tubular minima of the lattice potential into spatially well separated regions.

The experiments begin with a nearly pure Bose-Einstein condensate (BEC) of $^{87}$Rb atoms prepared in the $F=2, m_F=2$ hyperfine component of the ground state in a spherical harmonic magnetic trap with 40~Hz trap frequency. The wavelengths of the lattice potential (1064 nm) and the light sheet (763 nm) are detuned to the negative and positive sides of the relevant principle transitions of $^{87}$Rb at 780~nm and 795~nm, respectively, such that the lattice confines the atoms in the intensity maxima, while the light sheet produces a repulsive potential barrier. Initially, the BEC is adiabatically loaded into the lowest Bloch band. With the light sheet activated, both regions of the lattice above and below the light sheet (cf. Fig.~\ref{fig:2}(a)) are loaded in equal shares. As has been described in previous work \cite{Wir:11, Oel:13}, diabatic Landau-Zener sweeps of $\Delta V$ allow us to excite the atoms to the second Bloch band, which, in a certain range of $\Delta V$, is well protected against collisional relaxation. After applying the excitation sweep of $\Delta V$, the second band is nearly evenly populated. However, within the first 10~ms a substantial fraction of the population relaxes to its energy minima via collisions and tunneling processes. Collisional relaxation depletes the band only after several 100~ms. The condensation process in both regions is completely independent, since tunneling between the two sectors is suppressed by the presence of the light sheet. In Fig.~\ref{fig:2}(b), a band mapping technique was used to image the atomic populations in quasi-momentum space after giving the ensemble about 80~ms time to equilibrate. The line of sight in this image is the $z$-axis and the light sheet was switched off, however, no notable difference is observed for this viewing direction, if the light-sheet is activated. The observed quasi-momentum distribution shows a simultaneous approximately equal macroscopic occupation of both band minima. Condensate fractions of more than $30\%$ are found. Due to the elongate shape of the lattice sites, the interaction energy per particle is about two orders of magnitude lower than the tunneling energy, and hence the associated condensate states $|\psi_{\pm} \rangle$ are well approximated by Bloch states. 

\begin{figure*}[hbt]
\includegraphics[scale=1, angle=0, origin=c]{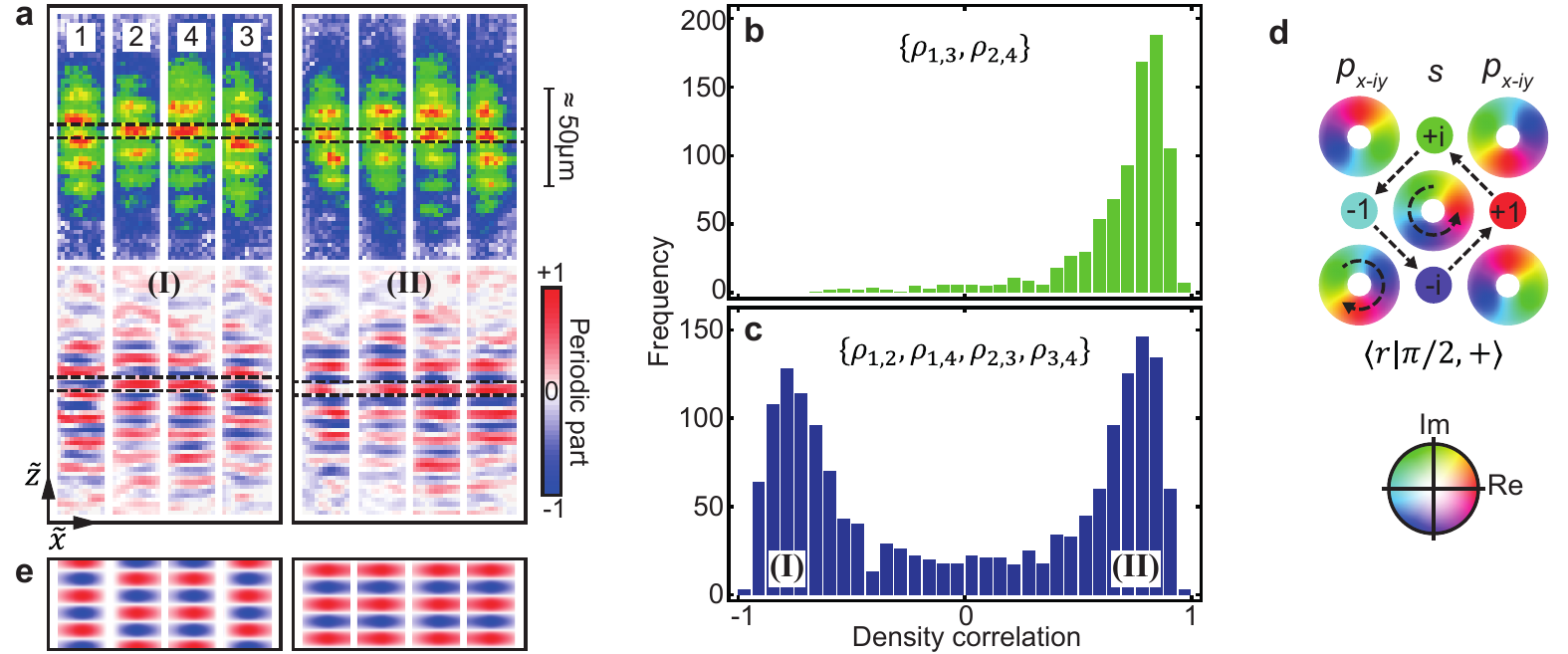}
\caption{\label{Fig.3} (color online). (a) The left (I) and right (II) panels show examples of the two classes of observed interference patterns. The upper row shows the original data, which are repeated in the lower row after low pass filtering and normalization. A data set comprises four density gratings $n_i(\tilde{z})$ with $i\in \{1,2,3,4\}$ corresponding to the four Bragg maxima in Fig.~\ref{fig:2}(c) and (d) (see text for definition of $n_i(\tilde{z})$). The black dashed boxes highlight the spatial correlations among the four density gratings. (b) Stacked histogram for the density-density cross correlation $\rho_{i,j}$ with both indices odd or even. (c) Stacked histogram for the density-density cross correlation $\rho_{i,j}$ with mixed indices. (d) Wave function associated with the state $|\pi/2,+\rangle$. The colors parametrize the local phase factor ranging from +1 (red), via +i (green) and -1 (cyan) to -i (purple). (e) A section of the calculated interference patterns is shown, which exhibit the same spatial correlations as found in (a).}
\label{fig:3}
\end{figure*}

The experimental protocol is illustrated in Fig.~\ref{fig:2}. In presence of the light sheet (a), condensates are simultaneously prepared in the $X_\pm$-points of the second band in two spatially separated regions of the lattice. Subsequently all potentials are switched off. After a ballistic expansion of $t_{\textrm{tof}} = 25\,$ms, which approximates a Fourier transform into momentum space, an absorption image is recorded along the $\bf{r}$-direction indicated in (c). The figure also illustrates the four zero-order Bragg peaks (indicated 1,2,3,4) observed at the four quasi-momenta in the $xy$-plane, associated with the condensation points $X_\pm$, each carrying an interference pattern due to the overlapping contributions from the condensates in both lattice regions. The main signature of the interference in each Bragg peak is a spatial modulation of the particle density along the $z$-direction. In the simplified picture that the interference results from two pulsed point sources at a distance $d_z$ observed after sufficiently long expansion time of flight $t_{\textrm{TOF}}$, the wavelength of the expected density grating is $\lambda_z = \frac{h t_{\textrm{TOF}}}{m d_z}$, with $h$ denoting Planck's constant and m the atomic mass \cite{And:97}. This amounts to $\lambda_z \approx 10\,\mu$m, close to what is observed in the experiment. An example of a single realization is shown in Fig.~\ref{fig:2}(d). In this image, the Bragg peaks are recorded along a line of sight within the $xy$-plane enclosing an angle of $13^\circ$ with the $y$-axis, defining the new coordinates $\tilde{x},\tilde{z}$ of the projected density distribution.

While the global spatial phases $\theta_i$ of the density gratings $i \in \{1,2,3,4\}$ in Fig.~\ref{fig:2}(c,d) vary for different experimental realizations, thus reflecting the complete independence of the atomic samples produced in the two regions of the lattice \cite{Jav:96}, the relative spatial phases $\theta_{i,j} \equiv \theta_i-\theta_j$ carry information about the underlying quantum state. For a quantitative correlation analysis of the density gratings, we perform $420$ independent experimental realizations, each yielding a set of four sectors (one for each of the 4 Bragg peaks in Fig.~\ref{fig:2}(c) and (d)) as is illustrated in the upper row of Fig.~\ref{fig:3}(a). Since we are only interested in the spatial phases of the interference gratings along the $z$-direction, we extract their periodic parts by removing the constant offset by low pass filtering (see Ref.~\cite{SM}) and get the distributions $n_{i}(\tilde{x},\tilde{z})$ shown in the lower row of Fig.~\ref{fig:3}(a). These distributions take values in the interval $[-1,1]$ and are centered around zero: $\sum_{\tilde{x},\tilde{z}} n_i(\tilde{x},\tilde{z})\approx 0$. After summing along the pixel values in the $\tilde{x}$-direction in order to obtain $n_i(\tilde{z}) \equiv \sum_{\tilde{x}} n_i(\tilde{x},\tilde{z})$, we evaluate the normalized density-density correlations
\begin{eqnarray}
\rho_{i,j} 	 \equiv \frac{\sum_{\tilde{z}} n_i(\tilde{z})n_j(\tilde{z})}{\sqrt{\sum_{\tilde{z}} n_i^2(\tilde{z}) \sum_{\tilde{z}} n_j^2(\tilde{z})}}.
\label{eq:}
\end{eqnarray}
Note that the value of the correlations $\rho_{i,j}$ is connected to the relative spatial phases between the interference patterns such that $\rho_{i,j}=+1$ if $\theta_{i,j} = 0$ and $\rho_{i,j} = -1$ if $\theta_{i,j} =\pi$. In Fig.~\ref{fig:3}(b) a histogram for the observed values of $\rho_{1,3}$ and $\rho_{2,4}$ is plotted, i.e. both indices are either odd or even. A clear accumulation point is observed near $+1$. Fig.~\ref{fig:3}(c) displays the histogram of all correlations $\rho_{i,j}$ with one odd and one even index. In contrast to (b), a bimodal distribution with practically equal numbers of correlated and anti-correlated patterns is observed. Examples of an anti-correlated (I) and a correlated (II) realization are shown in the left and right panels of Fig.~\ref{fig:3}(a). This central finding of our work leads to a remarkable consequence: consider each atomic subsample in the upper (u) and lower (l) regions of the lattice, indicated by $\nu \in \{u,l\}$ as a superposition $|\psi_{+}\rangle + e^{i\phi_{\nu}} |\psi_{-}\rangle$ with arbitrary and possibly indeterminate phases $\phi_{\nu}$, where "indeterminate" means that these phases arbitrarily change for different implementations. The single accumulation point in Fig.~\ref{fig:3}(b) reflects the fact that the Bragg peaks with odd or even indices belong to the same condensation point. Hence, the spatial phases of their interference gratings are necessarily always the same and we cannot learn anything from these correlations on the phases $\phi_{\nu}$. The bimodal distribution in Fig.~\ref{fig:3}(c), however, observed for Bragg peaks belonging to different condensation points, shows that the arbitrary phases $\phi_{\nu}$ are constrained by the condition $\phi_{u} - \phi_{l} \in \{0,\pi\}$ for all implementations. This correlation between $\phi_{\nu}, \nu \in \{u,l\}$, observed although both samples are completely independent, implies that $\phi_{\nu}$ must take specific values $\phi_0 \pm m_{\nu} \pi$ (with integer $m_{\nu}$), with the same inherent value of $\phi_0$ realized in each experimental realization. We conclude that in each lattice region (above and below the light sheet barrier) either of the coherent superpositions $|\phi_0,\pm \rangle \equiv \left(|\psi_{+}\rangle \pm e^{i \phi_0} \,|\psi_{-} \rangle \right)/\sqrt{2}$ is realized. An incoherent mixture $\frac{1}{2} \left( | \psi_+ \rangle \langle \psi_+ | + | \psi_- \rangle \langle \psi_- | \right)$ or spatially separated domains of the substates $\left | \psi_\pm\right \rangle$ are hence excluded. 

\begin{figure}
\includegraphics[width=0.45 \textwidth]{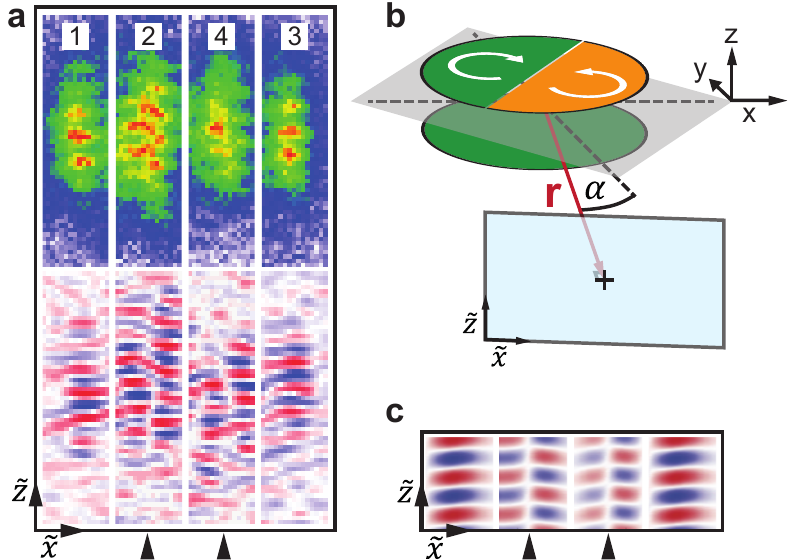}
\caption{\label{Fig.4} (color online). (a) Interference pattern of a slightly excited sample: the inner density gratings (2 and 4) show vertical nodal lines indicated by the arrows at the lower edge. (b) Schematic of the experimental scenario, which explains the observations. The atomic sample in the upper lattice region is divided into two equally sized domains of different chirality indicated by the orange and green areas. The sample in the lower lattice region(green area) carries no excitations. (c) Calculation of the interference based on the scenario illustrated in (b). The nodal lines corresponding to the observations are indicated by black arrows.}
\label{fig:4}
\end{figure}

It appears natural to expect that the emerging value of $\phi_0$ should minimize the free energy for the family of states $|\phi_0,\pm \rangle$. A detailed calculation in Ref.~\cite{Oel:13} (see also \cite{Liu:06}) shows that this applies for $\phi_0 = \pi/2$ for the following reason. While the states $|\phi_0,\pm \rangle$ have the same kinetic energy, irrespective of the value of $\phi_0$, the interaction energy explicitly depends on $\phi_0$. In the deep wells of the lattice potential, the local wave function is the superposition $p_x \pm e^{i \phi_0} p_y$, where $p_x$ and $p_y$ denote the wavefunctions of the local $p_x$ and $p_y$-orbitals, respectively. In analogy to Hund's second rule in multi-electron atoms, the interaction energy is minimized if $\phi_0 = \pi/2$, because the superposition $p_x \pm i p_y$ provides the largest possible mode volume such that the repulsively interacting atoms can best avoid each other. Note that Ref.~\cite{Oel:13} also presents experimental evidence for the minimization of interaction energy. We hence conclude that in each lattice region and in each experimental implementation either of the two possible minimal energy states $|\pi/2,+\rangle$ or $|\pi/2,-\rangle$ is produced via a spontaneous breaking of the inherent $Z_2$ symmetry associated with two possible signs of the chiral order. The corresponding wavefunction $\langle r |\pi/2,+\rangle$ is illustrated in Fig.~\ref{fig:3}(d). In the deep wells, it possesses alternating local $p_x \pm i p_y$ orbitals, thus maximizing angular momentum in each plaquette. The local phases are arranged to introduce plaquette currents breaking time-reversal symmetry and the translation symmetry of the lattice. Our findings are supported by detailed calculations. In Fig.~\ref{fig:3}(e) we show the interference contrast calculated for the two cases of equal condensates corresponding to states $|\pi/2,+ \rangle$ in both lattice regions (right panel) and for different condensates  $|\pi/2,+\rangle$ and $|\pi/2,- \rangle$ in the upper and lower lattice region, respectively (left panel). These pictures reproduce the signatures of the corresponding data in Fig.~\ref{fig:3}(a). Details are deferred to Ref.~\cite{SM}.
 
At higher temperatures, the correlations $\rho_{i,j}$ reduce and eventually, the distributions in the histograms in Fig.~\ref{fig:3}(b) and (c) wash out. Even at the lowest temperatures possible in our experiments some realizations give rise to distorted interference patterns with spatial irregularities, indicating the presence of excitations. Some specific low energy excitations (found in about $1 \%$ of all implementations) yield characteristic interference patterns, as shown in Fig. \ref{fig:4}(a). While the two outer interference gratings (1 and 3) are not notably different from those of Fig.~\ref{fig:3}(a), the inner gratings 2 and 4 show additional vertical nodal lines, highlighted by the two arrows. A zipper-like structure arises, more clearly seen in the high-pass filtered images shown in the bottom part of the graph. A theoretical analysis, deferred to Ref.~\cite{SM}, shows that this signature reflects the formation of two domains of different chirality in one of the lattice regions, while the other region does not carry notable excitations. These domains are nearly equally sized, and they are separated by a straight domain wall approximately aligned with the $x+y$-direction (see Fig. \ref{fig:4}(b)). In Fig. \ref{fig:4}(c), an interference pattern calculated for this specific configuration is plotted, which reproduces the signatures seen in the data in Fig. \ref{fig:4}(a) (see Ref.~\cite{SM}). The observed vertical nodal line arises from the destructive interference in the $x-y=0$-plane of the matter waves emanating from the two domains of the upper lattice region during ballistic expansion, which is expected due to their phase difference of $\pi$.  

\begin{acknowledgments}
This work was partially supported by DFG-SFB 925 and the Hamburg Centre for Ultrafast Imaging (CUI). We are grateful to Vincent Liu and Cristiane Morais Smith for useful discussions.
\end{acknowledgments}

\section{Supplemental Material}

\subsection{Theoretical analysis of interference patterns} 

\begin{figure}
\includegraphics[scale=0.2, angle=0, origin=c]{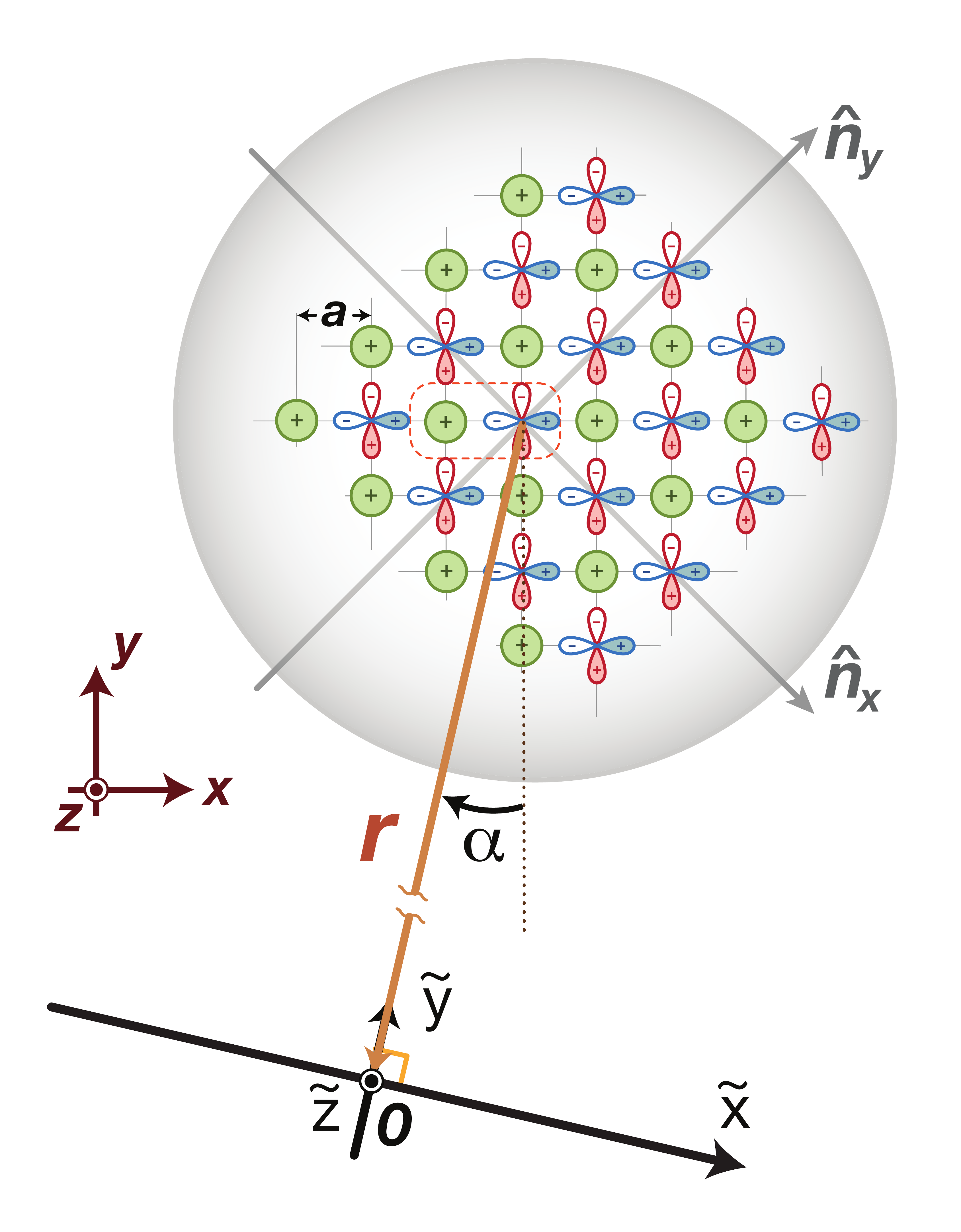}
\caption{\textbf{The coordinate setup for simulations.} The interference contrast is calculated in the $\tilde{x}\tilde{z}$-plane at a distance $|\bm{r}|$ from the lattice. The angle $\alpha\simeq 13^0$ is chosen according to the experimental setup.}
\label{2Dsection}
\end{figure}

The interference contrast of the time-of-flight images (TOF) corresponds to the spatially oscillatory part of the density-density correlation function $\left\langle\rho(\bm{R};t)\rho(\bm{R}+\tilde{\bm{R}};t)\right\rangle$ at time $t$ after the release from the trap. The density operator is $\rho(\bm{R};t)=\psi^{\dag}(\bm{R};t)\psi(\bm{R};t)$, and $\tilde{\bm{R}}$ describes the spatial distance of the two operators. The trapped system consists of one-dimensional tubes along the $z$-axis, which are centrally split by a barrier potential located within the $xy$-plane at $z=0$. The orbitals in the $xy$-plane are  Wannier states with $s$-, $p_{x}$- or $p_{y}$ geometry. They are arranged as shown in Fig.~\ref{2Dsection}. We approximate these Wannier states by the eigenstates of a harmonic oscillator:
\begin{eqnarray}
W_{\nu}(\bm{r}-\bm{n})\hspace{-0.1cm}=\hspace{-0.1cm}\frac{c_{\nu}}{\sqrt{\pi}l} e^{-\frac{1}{2l^2}\left|\bm{r}-\bm{n}\right|^2},
\end{eqnarray}
with the oscillator length $l$ and $\bm{r}\equiv(x,y)$. The coefficients are defined as $c_1=1$ and $c_{2/3}=\left[(x-n_x+\frac{a}{\sqrt{2}})\pm(y-n_y+\frac{a}{\sqrt{2}})\right]/l$ with the lattice constant $a$. The Wannier orbitals are centered at the lattice locations $\bm{n}$, as depicted in Fig.~\ref{2Dsection}. These orbitals expand ballistically during TOF, resulting in time-dependent orbitals $W_{\nu}(\bm{r}-\bm{n}, t)$, given below.

In the $z$-direction, the expansion of the one-dimensional clouds is more involved. We approximate this expansion with a single, ballistically expanding orbital. At time $t=0$ it is the ground state of a harmonic oscillator, with a harmonic oscillator length scale $d_z$. This scale and the corresponding energy $\hbar^{2} /(2 m d_z^{2})$ is the effective, typical energy scale of the atomic expansion in the z-direction, which consists of both the initial kinetic energy in z-direction, and the interaction energy that has been converted into kinetic energy in the early stage of the  TOF expansion. The length scale $d_z$ controls the scale of the interference fringes and could be determined  experimentally from these. With this approximation, the  time evolution of the field operator can be separated as  $\psi\left(\bm{R};t\right)=T_d(z;t)\psi_{d}\left(\bm{r};t\right)+T_u(z;t)\psi_{u}\left(\bm{r};t\right)$, see ~\cite{Pol:06}. The propagator along the $z$-direction is $T_j(z;t)=e^{iq_{j}z-iq_j^2 t/2m}$ with $q_{u,d}=m(z\pm d_z/2)/\hbar t$, and $j= u, d$ indicating the upper and lower half-spaces with $z>0$ and $z<0$, respectively. The projection of the interference contrast  along the imaging direction $\tilde{y}$ (see Fig.~\ref{2Dsection}) is 
\begin{eqnarray}\label{IF1}
&&\nonumber \hspace{-1.3cm}I\left(\bm{r},\tilde{x},\tilde{z}\right)=\frac{1}{L}\int_0^L d\tilde{y} \Big\langle\rho(\bm{R};t)\rho(\bm{R}+\tilde{\bm{R}};t)\Big\rangle_{\textrm{int}} \\
&&\hspace{0.75cm} = \frac{1}{L}\left[\mathcal{A}_q\left(\bm{r},\tilde{x};t\right)e^{iq\tilde{z}}+{\rm h.c.}\right]\, .
\end{eqnarray}

Here, the density-density correlation $\Big\langle\rho(\bm{R};t)\rho(\bm{R}+\tilde{\bm{R}};t)\Big\rangle$ is written as a sum of a term slowly varying with respect to $\tilde{z}$, which is irrelevant for the interference, and a term $\Big\langle\rho(\bm{R};t)\rho(\bm{R}+\tilde{\bm{R}};t)\Big\rangle_{\textrm{int}}$ that rapidly oscillates with $\tilde{z}$ and essentially comprises the product of the in-situ Green's functions of both subsytems: 
\begin{eqnarray}\label{IF2}
\mathcal{A}_q\left(\bm{r},\tilde{x};t\right)\hspace{-0.05cm}=\hspace{-0.15cm}\int_0^L d\tilde{y}\hspace{0.1cm}\mathcal{G}_{u}(\bm{r},\bm{r}+\tilde{\bm{r}};t) \,\mathcal{G}^*_{d}(\bm{r},\bm{r}+\tilde{\bm{r}};t),
\end{eqnarray}
with $q=md_z/\hbar t$ and the imaging length $L$. We relate the field operators $\psi_{j}\left(\bm{r};t\right)$ to the lattice operators $b_{j\nu}(\bm{n})$ via 
 
\begin{eqnarray}
&&\hspace{-1.0cm}\psi_{j}\left(\bm{r};t\right)=\sum_{\nu}\sum_{\bm{n}}W_{\nu}(\bm{r}-\bm{n};t)b_{j\nu}(\bm{n}).
\end{eqnarray}

With this, the interference contrast is entirely controlled by the single particle correlation functions  $\langle b^{\dagger}_{j\nu}(\bm{n})b_{j\mu}(\bm{m})\rangle$ in each subsystem. In the following, we calculate these for condensates with chiral $p_x \pm ip_y$ symmetry. We thus obtain Fig.~3(e) of the main text, where we plot the interference contrast for the case of two sub-systems with opposite chirality (left panel) and with the same chirality (right panel). Furthermore, we can determine the interference signature in presence of defects, by choosing the in-situ correlation functions accordingly. In Fig.~4(c) of the main text we show the interference pattern of two chiral BECs, when one of them has a domain wall defect sketched in Fig.~4(b). 

\textbf{The bosonic Green's function} As pointed out above, the interference patterns are related to the two-dimensional bosonic Green's function $\mathcal{G}_{j}(\bm{r}_1,\bm{r}_2;t)$. Adopting the coordinate conventions illustrated in Fig.~\ref{2Dsection}, we expand the bosonic fields in the Wannier basis and Fourier transform to momentum space 
\begin{eqnarray}
&&\hspace{-1.0cm}\psi_{j}\left(\bm{r};t\right)=\sum_{\nu}\sum_{\bm{n}}W_{\nu}(\bm{r}-\bm{n};t)b_{j\nu}(\bm{n}),\\
\nonumber
&&\hspace{ -0.1cm}=\frac{1}{\sqrt{N}}\sum_{\nu}\sum_{\bm{n}}\sum_{\bm{k}}W_{\nu}(\bm{r}-\bm{n};t)\hspace{0.1cm}e^{i\bm{k}\cdot\bm{n}}b_{j\nu}(\bm{k}),
\end{eqnarray}
where $\nu=1,2,3$ denote the three different orbitals within one unit cell, and $\bm{n}=n_x\hat{n}_x+n_y\hat{n}_y$ with $n_{x(y)}/\sqrt{2}a=-\left(N-1\right)/2,\cdots,-1/2,1/2,\cdots,\left(N-1\right)/2$. We employ Gaussian-like time-dependent Wannier functions
\begin{eqnarray}
W_{\nu}(\bm{r}-\bm{n};t)\hspace{-0.1cm}=\hspace{-0.1cm}\frac{\mathcal{C}_{\nu}}{\sqrt{\pi}l_t}e^{-i\theta_t}e^{-\frac{1}{2l_t^2}\left|\bm{r}-\bm{n}\right|^2}e^{i\left(\frac{\hbar t}{ml^2}\right)\frac{1}{2l_t^2}\left|\bm{r}-\bm{n}\right|^2},
\end{eqnarray}
where the time-dependent width is defined as $l_t=l\sqrt{1+\left(\frac{\hbar t}{ml^2}\right)^2}$ with the initial width $l$ at $t=0$, $\theta_t=\tan^{-1}\left(\frac{\hbar t}{ml^2}\right)$, and the coefficients $\mathcal{C}_1=1$ and $\mathcal{C}_{2/3} = \,c_{2/3} \, e^{-i\theta_t}\, l /l_t$. We note that $W_{1}(\bm{r}-\bm{n};t)$ represents the isotropic $s$-orbitals sketched by the green disks in Fig.~\ref{2Dsection}. The Wannier functions $W_{2}(\bm{r}-\bm{n};t)$ and $W_{3}(\bm{r}-\bm{n};t)$ represent the $p_y$- and $p_x$-orbitals, illustrated as blue and red bar-bells in Fig.~\ref{2Dsection}. Diagonalizing the tight-binding Hamiltonian in momentum space~\cite{Oel:13} yields the band representation $b_{j\nu}(\bm{k})=\sum_{a}M_{\nu a}(\bm{k})\tilde{b}_{ja}(\bm{k})$ with the transformation matrix $M_{\nu a}$. Since at low temperature only the lowest energy band is occupied, one obtains $b_{j\nu}(\bm{k})\simeq M_{\nu 1}(\bm{k}) \tilde{b}_{j1}(\bm{k})$. Simplifying notation by writing $b_j \equiv \tilde{b}_{j1}$ the Green's function reads
\begin{eqnarray}
\label{greenN}
\nonumber&&\hspace{-.8cm}\mathcal{G}_{j}(\bm{r}_1,\bm{r}_2;t)\\
&&\nonumber\hspace{-0.5cm}\simeq\frac{1}{N^2}\sum_{\nu\mu}\sum_{\bm{n},\bm{m}}\sum_{\bm{k}_1,\bm{k}_2}W_{\mu}(\bm{r}_2-\bm{m};t)W_{\nu}^*(\bm{r}_1-\bm{n};t)\\
&&\hspace{0.cm}e^{-i\bm{k}_1\cdot\bm{n}}e^{i\bm{k}_2\cdot\bm{m}}M_{\mu 1}(\bm{k}_2)M_{\nu 1}^*(\bm{k}_1)\Big\langle b^{\dag}_j(\bm{k}_1) b_j(\bm{k}_2)\Big\rangle.
\end{eqnarray}
For the numerical simulation the hopping parameters used to compute the transformation matrix for the numerical simulation are taken from Ref.~\cite{Oel:13}.

\textbf{Computation of the interference patterns} Collisional interaction acts to scatter boson pairs among the two energy minima of the lowest-energy band such that superposition states involving both minima are energetically favored (cf. Ref.~\cite{Oel:13}). Thus the ground states can be represented in a binomial form with respect the two energy minima,
\begin{eqnarray}
\label{GroundState}
\left|\Psi_j\right\rangle_{\theta,\phi}=\frac{1}{\sqrt{N_0!}}\left[\cos(\theta_j) \,b_{j+}^{\dag}+e^{i\phi_j}\sin(\theta_j) \,b_{j-}^{\dag}\right]^{N_0}\left|0,0\right\rangle,
\end{eqnarray}
where $b_{j\pm} \equiv b_j(\bm{k}_{\pm})$, $\bm{k}_{\pm} \equiv  (1\mp1,1\pm1 )\,\pi / (\sqrt{8}\,a)$ denotes the quasi-momenta at the two minima, and $N_0$ is the total number of bosons. In this expression, $\theta_j$ determines the relative number of bosons in the two minima, and $\phi_j$ represents the phase difference between the bosons in the two minima. At zero temperature, the ensemble average $\Big\langle b^{\dag}_j(\bm{k}_1)b_j(\bm{k}_2)\Big\rangle$ in Eq.~(\ref{greenN}) results as the expectation value with respect to the ground state of Eq.~(\ref{GroundState}) 
\begin{eqnarray}
&&\hspace{-0.8cm}\nonumber\frac{1}{N_0}\Big\langle b^{\dag}_j(\bm{k}_1)b_j(\bm{k}_2)\Big\rangle \\
&&\hspace{-0.8cm}\nonumber=\cos^2(\theta_j)\,\delta_{\bm{k}_1,\bm{k}_+}\delta_{\bm{k}_2,\bm{k}_+}+\sin^2(\theta_j)\,\delta_{\bm{k}_1,\bm{k}_-}\delta_{\bm{k}_2,\bm{k}_-}\\
&&\hspace{-0.9cm}+\sin(\theta_j)\cos(\theta_j)\,\left[e^{i\phi_j}\delta_{\bm{k}_1,\bm{k}_+}\delta_{\bm{k}_2,\bm{k}_-}+e^{-i\phi_j}\delta_{\bm{k}_1,\bm{k}_-}\delta_{\bm{k}_2,\bm{k}_+}\right].
\end{eqnarray}
In the weakly repulsive regime, relative phases $\phi_j = \pm \pi/2$ are energetically preferred (see Ref.~\cite{Oel:13}) yielding a $p_x\pm ip_y$ ground state with broken time-reversal symmetry. In our experiment the boson numbers of the two minima are almost equal. Thus we set $\theta_j=\pi/4$ and $\phi_j=\pm \pi/2$ in our numerical simulations. Assuming equal populations in the upper and lower lattices and choosing $l=0.3\,a$, $|\bm{r}|\simeq130\,a$, $L\simeq 2|\bm{r}|$, $\alpha=13^0$, $N=20$, $\frac{\hbar t}{m\,a^2} = 37$, we first compute the Green's function at zero temperature. Then, by using Eq.~(\ref{IF1}), we construct the interference contrast for atomic samples in both lattice sections exhibiting equal or opposite chirality. This yields the interference patterns plotted in Fig.~3(d) of the main text.

\textbf{Simulation for two domains in the upper lattice} When the upper lattice section hosts two domains with opposite chirality, the Green's function can be represented as 
\begin{eqnarray}
&&\hspace{-0.7cm}\nonumber \left \langle \psi_{u}^{\dag}(\bm{r}_1;t) \psi_{u}(\bm{r}_2;t) \right \rangle \\ 
&&\nonumber\hspace{-0.6cm}=\hspace{-0.3cm} \sum_{n=0,\pm1} \hspace{-0.3cm}P_{n} \left \langle \Phi_{N_1+n},\Phi_{N_2-n}\right|\psi_{u}^{\dag}(\bm{r}_1;t)\psi_{u}(\bm{r}_2;t)\left|\Phi_{N_1},\Phi_{N_2}\right\rangle.\\
\end{eqnarray}
Here, $\left|\Phi_{N_1},\Phi_{N_2}\right\rangle$ denotes the quantum state with $N_1$ and $N_2$ bosons condensed in the two domains, respectively, and $P_n$ is the probability of exchanging $n$ bosons between the two domains.  Upon the simplifying assumption that the particle number in each domain is fixed, the two dimensional Green's function of the upper lattice is approximated as
\begin{eqnarray}
&&\hspace{-0.9cm}\nonumber
\left\langle\psi_{u}^{\dag}(\bm{r}_1;t)\psi_{u}(\bm{r}_2;t)\right\rangle \\ 
&&\nonumber
\hspace{-0.7cm}\simeq\left\langle \Phi_{N_1},\Phi_{N_2}\right|\psi_{u}^{\dag}(\bm{r}_1;t) \psi_{u}(\bm{r}_2;t) \left|\Phi_{N_1}, \Phi_{N_2}\right\rangle \\
&&\hspace{-0.7cm}=\left\langle\psi_{u}^{\dag}(\bm{r}_1;t)\psi_{u}(\bm{r}_2;t)\right\rangle_{N_1}\hspace{-0.2cm}+\left\langle\psi_{u}^{\dag}(\bm{r}_1;t)\psi_{u}(\bm{r}_2;t)\right\rangle_{N_2},
\end{eqnarray}
where $\left\langle\psi_{u}^{\dag}(\bm{r}_1;t)\psi_{u}(\bm{r}_2;t)\right\rangle_{N_i}$ describes the propagator of the $i$-th domain for $N_i$ bosons. This yields the interference patterns plotted in Fig.~4(c) of the main text, calculated for a domain wall at $n_x=0$.

\subsection{Lattice set-up}
The optical square lattice potential provides two different lattice sites per unit cell denoted as A and B. They are arranged as the black and white fields of a chequerboard. The well depth difference between these two sites $\Delta V\equiv V_A-V_B$ and the average well depth $V_0$ can be dynamically adjusted. The potential is given by
\begin{equation}
\label{potential}
V(x, y) \equiv -\frac{V_0}{4} \left|\eta \left(e^{ikx}+\epsilon_x e^{-ikx} \right)+ e^{i\theta}\left(e^{iky}+\epsilon_y e^{-iky} \right)\right|^2,
\end{equation}
with the laser wavelength $\lambda=2\pi/k=1064$~nm, the time-phase difference $\theta$ between the two oscillating standing waves and the asymmetry parameters $\epsilon_x,\epsilon_y,\eta$ which account for an intensity imbalance between all four involved laser beams. The active stabilization of $\theta$ by means of an interferometric set-up (with a precision better than $\pi/300$) leads to a direct control of $\Delta V = V_0(1+\epsilon_x)(1+\epsilon_y) \cos (\theta)$. To ensure the degeneracy of the states at $X_\pm$ in the second Bloch-band we set the asymmetry parameter in the $y$-direction to $\epsilon_y=1$ \cite{Oel:13}. The other parameters are given by $\epsilon_x=0.81, \eta=1.11$. The lattice system is superimposed with a spherical harmonic magnetic trap with 40~Hz trap frequency. This leads to tubular shaped lattice sites elongated in the $z$-direction.

\begin{figure}
\includegraphics[scale=0.9, angle=0, origin=c]{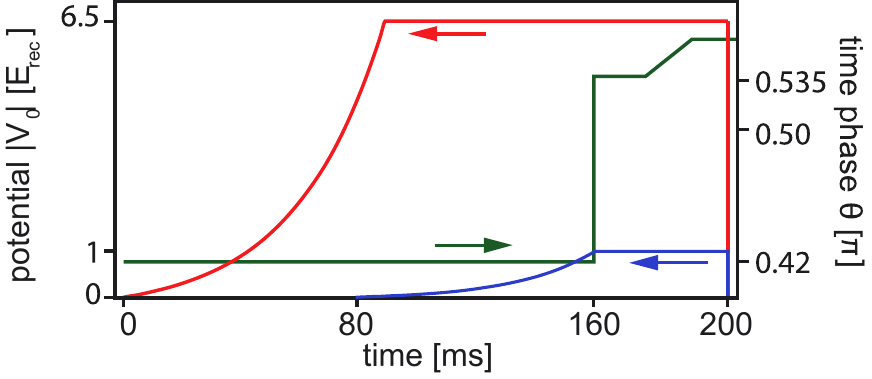}
\caption{\textbf{Experimental sequence}. The optical lattice is ramped to $\approx 6.5\ ~E_{\rm rec}$ depth within 80~ms (red trace). Within the subsequent 80~ms, the potential barrier is ramped up (blue trace). Finally, the second band is loaded of via tuning of $\theta$ in Eq.~(\ref{potential}) (green trace).}
\label{SI-RampUp}
\end{figure}

\begin{figure*}[hbt]
\includegraphics[scale=1, angle=0, origin=c]{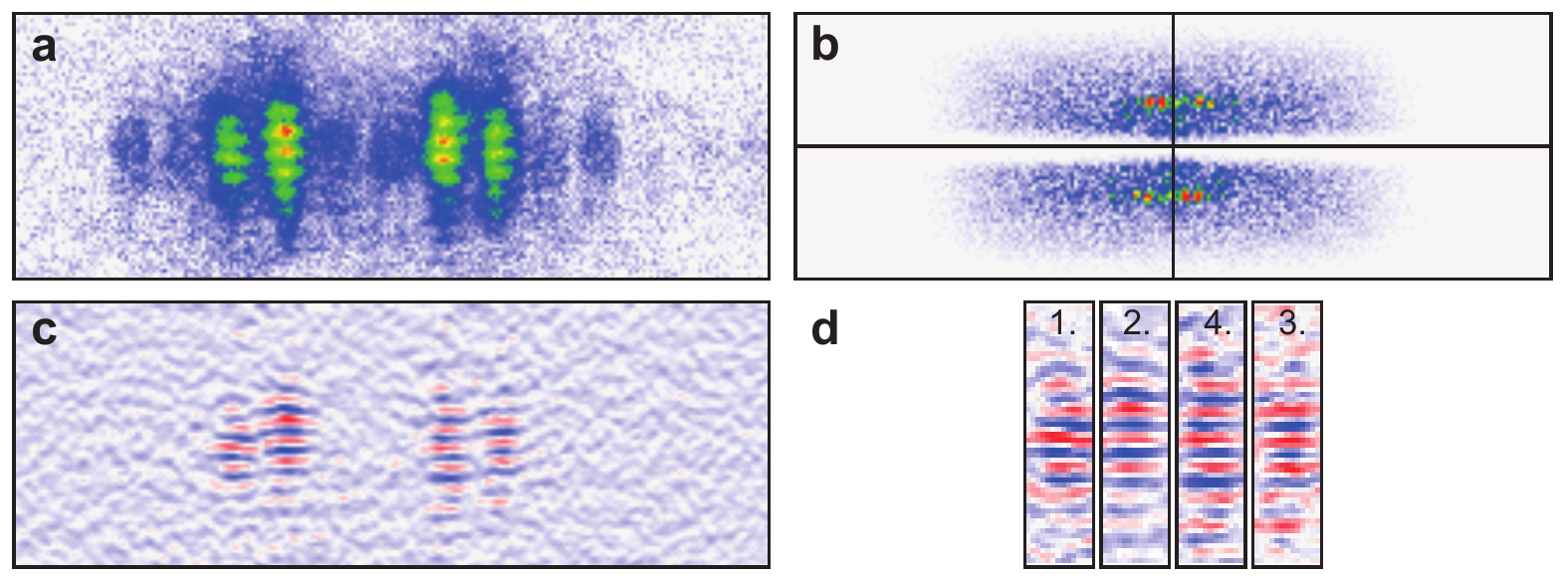}
\caption{\textbf{Data analysis}. Each image (a) is Fourier transformed and multiplied with generalized normal window functions of second order to yield (b). After an inverse Fourier transform (c) is obtained. Combining sections each comprising one of the interference patterns in (c) yields (d), which is used for further analysis.}
\label{Fig_S3}
\end{figure*}

\subsection{Experimental Procedure}
Our experiment starts with the preparation of a nearly pure Bose-Einstein condensate (BEC) ($^{87}$Rb, $N\approx50\cdot10^3$, $F=2,m_F=2$) in the magnetic trap (40~Hz trap frequency) at a temperature at least a factor 4 below the critical temperature. Subsequently we apply the experimental sequence sketched in Fig.~\ref{SI-RampUp}. Within the first 80~ms the BEC is adiabatically loaded into the lattice potential with a time-phase difference set to $\theta=0.42~\pi$ and a final depth of $V_0=6.5 ~E_\text{rec}$. The atoms are located in the deeper B-sites and the tunneling rates are on a low level resulting in an almost evenly populated first Bloch-band. In the next step the system gets separated into two identical layers by a potential barrier applied in the $xy$-plane. This barrier is provided by a laser beam with a wavelength of 763~nm, which is blue detuned with respect to the relevant atomic transitions of Rb. The beam propagates in the $x$-direction and the beam waist radius is elliptically shaped with $w_y=32~\mu$m and $w_z=4.5~\mu$m. A slow increase of the barrier within 80~ms to a final height of $2~E_\text{rec}$ ensures a controlled separation of the two clouds by approximately 10~$\mu$m. At this stage the two systems evolve completely independently since the particle exchange is suppressed. 

To excite the two ensembles into the second Bloch-band we change the sign of $\Delta V$ by setting the time-phase difference to $\theta=0.535~\pi$ within $200~\mu$s. After the excitation procedure we wait a total of 40~ms to let the system equilibrate. This duration is longer than the condensation process to the two minima of the band (10~ms) and shorter than the lifetime due to interband relaxation (200~ms). During this period we adiabatically increase the population in the $p$-orbitals by setting the time-phase difference to $\theta=0.56~\pi$ to suppress fluctuations of the population difference between the two condensation points \cite{Oel:13}. Finally, all potentials are switched off, the two ensembles expand ballistically and get imaged after 25~ms. 

\subsection{Data analysis}
The optical axis of the imaging system $\tilde{y}$ runs parallel with respect to the $xy$-plane. We introduced a small angle of $13^\circ$ between $\tilde{y}$ and the $y$-axis of the lattice system allowing to distinguish all zero order Bragg-resonances in the images. Each Bragg peak comprises a density grating along the $z$-axis due to interference from contributions from the two lattice regions above and below the potential barrier shown in Fig.~2(a) of the main text. Each density grating is a positive oscillating function with respect to the $z$-axis, which may be decomposed into a constant offset and an oscillatory part with vanishing integral. Since we are interested to determine the relative spatial positions of the density gratings shown in Fig.~3(a) of the main text, it appears natural to subtract the irrelevant offset. Hence, as a first step we extract the oscillating parts of these gratings by band-pass filtering. This procedure is illustrated in Fig.~\ref{Fig_S3}. We Fourier transform the original image Fig.~\ref{Fig_S3}(a) and use two generalized normal window functions of second order with different radii to block the DC offset and all frequencies much higher than the spatial frequency of the interference pattern. This leads to Fig.~\ref{Fig_S3}(b) and after a subsequent inverse Fourier transform to Fig.~\ref{Fig_S3}(c). We combine sections including each interference pattern of Fig.~\ref{Fig_S3}(c) to get Fig.~\ref{Fig_S3}(d), which corresponds to Fig.~3(b) of the main text. The optical axis of our imaging system $\tilde{y}$ is actually slightly tilted ($\approx 2^\circ$) out of the $xy$-plane. This introduces a small shift between the Bragg-resonances located in the front and rear object planes (cf. Fig.~2(c) of the main text). We account for this in all images by translating pattern 1 and 4 into the negative $z$-direction by an amount of two pixels with respect to pattern 2 and 3.

\end{document}